\documentclass[reprint, aps, pra, superscriptaddress]{revtex4-1}

\usepackage{graphicx}
\usepackage{bm}
\usepackage{amsmath}

\let\Oldeqref\eqref
\renewcommand{\eqref}[1]{\text{equation }\Oldeqref{#1}}
\newcommand{\Eqref}[1]{\text{Equation }\Oldeqref{#1}}

\newcommand{\xx}{\mathbf{r}}
\newcommand{\xa}{\xx_a}
\newcommand{\xb}{\xx_b}
\newcommand{\kk}{\mathbf{k}}

\newcommand{\ka}{\kk_a}
\newcommand{\kb}{\kk_b}
\newcommand{\Dx}{\Delta\xx}
\newcommand{\Dxa}{\Delta\xa}
\newcommand{\Dxb}{\Delta\xb}
\newcommand{\Dk}{\Delta\kk}
\newcommand{\Dka}{\Delta\ka}
\newcommand{\Dkb}{\Delta\kb}
\newcommand{\bkappa}{\bm{\kappa}}
\newcommand{\brho}{\bm{\rho}}
\begin{document}
	
	\title{The generalized optical memory effect}
	
	\author{Gerwin Osnabrugge$^*$,$^\dagger$}
	\affiliation{Biomedical Photonic Imaging Group, MIRA Institute for Biomedical Technology \& Technical Medicine, University of
		Twente, PO Box 217, 7500 AE Enschede, The Netherlands}
	\email[Corresponding author: ]{g.osnabrugge@utwente.nl}
	\author{Roarke Horstmeyer$^\dagger$}
	\affiliation{Bioimaging and Neurophotonics Lab, NeuroCure Cluster of Excellence, Charit\'e Berlin and Humboldt University, Berlin, Germany}
	\affiliation{Future address: Biomedical Engineering Department, Duke University, Durham, NC 27708, USA}
	\email[Both authors contributed equally]{}
	\author{Ioannis N. Papadopoulos}
	\affiliation{Bioimaging and Neurophotonics Lab, NeuroCure Cluster of Excellence, Charit\'e Berlin and Humboldt University, Berlin, Germany}
	\author{Benjamin Judkewitz}
	\affiliation{Bioimaging and Neurophotonics Lab, NeuroCure Cluster of Excellence, Charit\'e Berlin and Humboldt University, Berlin, Germany}
	\author{Ivo M. Vellekoop}
	\affiliation{Biomedical Photonic Imaging Group, MIRA Institute for Biomedical Technology \& Technical Medicine, University of
		Twente, PO Box 217, 7500 AE Enschede, The Netherlands}
	
	\date{\today}
	
	\begin{abstract}
		The optical memory effect is a well-known type of wave correlation that is observed in coherent fields that scatter through thin and diffusive materials, like biological tissue. It is a fundamental physical property of scattering media that can be harnessed for deep-tissue microscopy or 'through-the-wall' imaging applications. Here we show that the optical memory effect is a special case of a far more general class of wave correlation. Our new theoretical framework explains how waves remain correlated over both space and angle when they are jointly shifted and tilted inside scattering media of arbitrary geometry. We experimentally demonstrate the existence of such coupled correlations and describe how they can be used to optimize the scanning range in adaptive optics microscopes.
	\end{abstract}
	
	\maketitle
	
	\section{Introduction \label{sect:Intro}}
\begin{figure}
		\includegraphics[width=1\columnwidth]{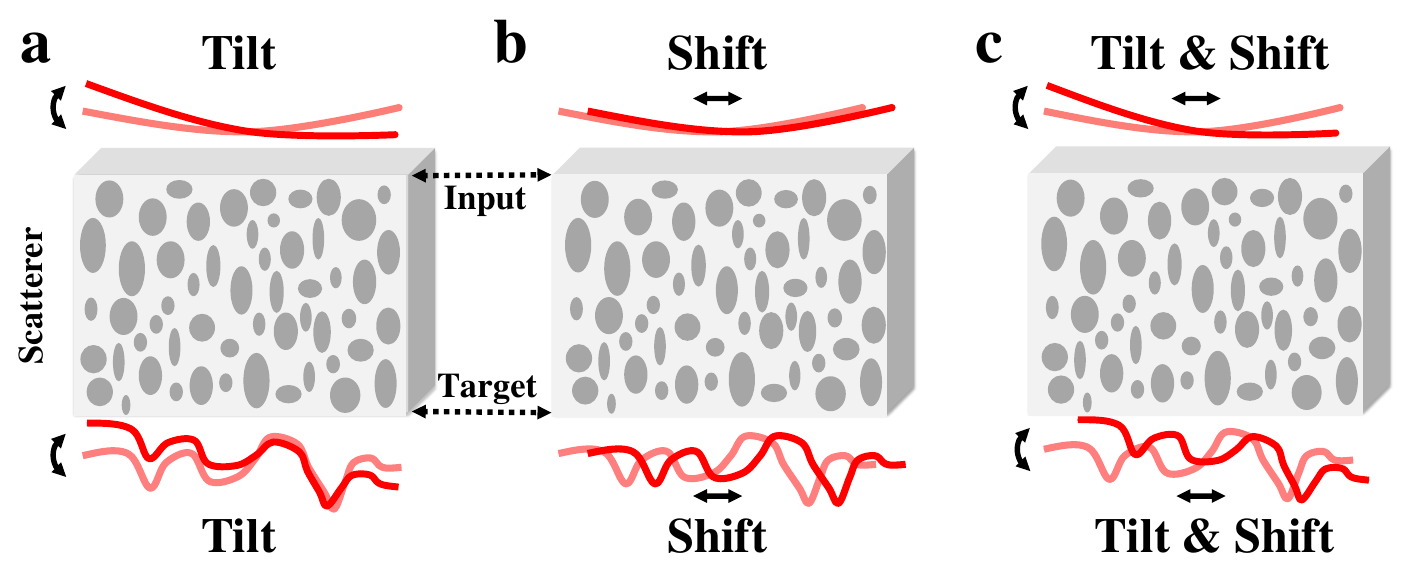}
		\caption{Three different types of spatial correlations in disordered media. (\textbf{a}) The optical `tilt' memory effect~\cite{Feng:1988}, where an input wavefront tilt leads to a tilt at the target plane. (\textbf{b}) The anisotropic 'shift' memory effect~\cite{Judkewitz:2015}, where an input wavefront shift also shifts the target plane wavefront. (\textbf{c}) Our new generalized memory effect, relying on both tilts and shifts, can maximize correlations along the target plane for a maximum imaging/focus scanning area.}
		\label{fig:memory-effects}
	\end{figure}

	It is challenging to record clear images from deep within biological tissue. As an optical field passes through tissue its spatial profile becomes randomly perturbed, which subsequently results in a blurry image of the features that lie underneath.  Luckily, even highly scattered optical fields still maintain a certain degree of correlation. Such scattering-based correlations have recently enabled new `hidden imaging' approaches ~\cite{Bertolotti:2012, Katz:2012, Katz:2014, Yang:2014}, which reconstruct clear images from behind diffusive materials. These prior investigations have primarily exploited what is traditionally referred to as the optical memory effect ~\cite{Freund:1988, Feng:1988}. This effect predicts that a scattered wavefront will tilt, but otherwise not change, when the beam incident upon a scattering material is also tilted by the same amount (see Fig.~\ref{fig:memory-effects}a)). These correlations have been observed through both thin isotropically scattering screens \cite{Li:1994} as well as thick forward scattering tissue~\cite{Schott2015}.
	
	Recently, we reported a new type of `shift' memory effect, illustrated in Fig.~\ref{fig:memory-effects}b, that occurs primarily in anisotropically scattering media~\cite{Judkewitz:2015}. This form of correlation is especially important in biomedical imaging, as it offers the ability to physically shift (as opposed to tilt) a focal spot formed deep \emph{within} scattering tissue by translating an incident optical beam.
	
	Here, we show how the optical `tilt' and `shift' memory effects are manifestations of a more general source of correlation within a scattering process, which depends upon how an incident wavefront is both tilted and shifted (see Fig.~\ref{fig:memory-effects}c). Our new `generalized memory effect' model offers a complete description of the first-order spatial correlations present within scattering media. While it applies to coherent waves in general, we limit our attention to the optical regime, where its applications range from maximizing the isoplanatic patch of adaptive optics microscopes~\cite{Mertz2015}
	to `hidden imaging' approaches~\cite{Bertolotti:2012, Katz:2012, Katz:2014, Yang:2014}. Here we develop and verify a general model that predicts an optimal imaging/scanning strategy for a given sample.
	We start out by presenting our model for the generalized memory effect. We describe how to predict the amount of expected correlation through a given slab of scattering material as function of both position and wavevector. Then, we experimentally measure the generalized memory effect and discuss the direct applications of our findings to adaptive optics.
	
	\section{Model \label{sect:Model}}
	In our model, we consider propagation of monochromatic and coherent light from an “input” plane $a$ to a “target” plane $b$. We limit ourselves to scalar waves. The forward-propagating field at the “input” surface, $E_a$, and at the “target” plane, $E_b$, are coupled through
	\begin{equation}\label{eq:TM-definition}
	E_b(\xb) = \int T(\xb, \xa) E_a(\xa) d^2\xa.
	\end{equation}
	It should be noted here that our model is completely general and does not make any assumptions about the location or orientation of the two planes, nor about the symmetry, shape, or other properties of the scattering sample. For example, for transmission through a slab of scattering material, the target plane is located at the back surface of the sample and $T(\xb, \xa)$ is the sample's transmission matrix. Conversely, in a biomedical microscope setup, the target plane is typically located \emph{inside} a scattering sample like tissue, and $T$ is a field propagator connecting position $\xa$ to a position $\xb$ located inside the tissue. 
	
	At this point let's define exactly what type of correlations we are interested in. Suppose that an input field $E_a$ results in a target field $E_b$. We hypothesize that when we shift/tilt the incident field with respect to the medium, the new target field will also experience a shift/tilt and also remain somewhat similar to the original target field. This similarity can be expressed as a correlation function involving the original matrix $T$ and a new matrix $\tilde{T}$ associated with the shift/tilted field. We describe `tilting' the incident wave as a multiplication with a phase ramp, $\exp{(i \Dka\cdot\xa)}$, resulting in a wavefront that is tilted by $\Dka$ with respect to plane $a$. Likewise, we describe the corresponding tilt at the target plane via a multiplication with $\exp{(-i \Dkb\cdot\xb)}$.
	
	We may now describe the new shifted/tilted situation by writing the corresponding shift/tilt matrix as $\tilde{T}(\xb,\xa)=\exp{(-i \Dkb\cdot\xb)} T(\xb+\Dxb,\xa+\Dxa) \exp{(i \Dka\cdot\xa)}$, where $\tilde{T}$ is shifted by $\Dxa$ and $\Dxb$ at the input and target plane, respectively. By calculating the ensemble averaged value $\langle T \tilde{T}^* \rangle$, we can find the corresponding correlation coefficient between the original fields (from $T$) and the tilted/shifted fields (from $\tilde{T}$). However, we found that the problem is expressed more naturally if we first introduce the center-difference coordinates $\xa^+\equiv\xa+\Dxa/2$, $\xa^-\equiv\xa-\Dxa/2$ (and the same for $\xb$). Using the same reasoning as above, we now find a symmetric expression for the field correlation function:
	\begin{equation}\label{eq:CW-definition}
	\begin{split}
	C_W(\Dxb, \Dkb; \Dxa, \Dka) \equiv \\
	\iint \langle 
	T(\xb^+, \xa^+) T^*(\xb^-, \xa^-)
	\rangle
	e^{i(\Dka\cdot\xa-\Dkb\cdot\xb)} d^2\xa d^2\xb,
	\end{split}
	\end{equation}
	In the special case that $\Dka=\Dkb$ and $\Dxa=\Dxb=0$, \eqref{eq:CW-definition} reduces to the optical `tilt' memory effect~\cite{Feng:1988}, whereas for $\Dxa = \Dxb$ and $\Dka=\Dkb=0$, the correlation function corresponds to the anisotropic `shift' memory effect described in Ref.~\cite{Judkewitz:2015}.
	
	Before proceeding to calculate $C_W$ in terms of the sample properties, we introduce the Wigner distribution function (WDF) of the optical field. The WDF describes an optical field as a joint `phase space' function of two variables: one spatial variable and one wavevector variable. We choose to work with the WDF because it will allow us to convert \eqref{eq:TM-definition} into a function of both space and wavevector, which we may easily connect to our new correlation function $C_W$ of similar variables. We define the WDF of field $E$ as
	\begin{equation}\label{eq:Wigner-definition}
	W(\xx,\kk) \equiv \int E(\xx^+) E^*(\xx^-) e^{-i\kk \cdot \Dx} d^2\Dx.
	\end{equation}
	While not always exact, it is convenient to connect the value $W(\xx,\kk)$ to the amount of optical power at point $\xx$ that is propagating in direction $\kk$~\cite{Bastiaans:79}.  
	
	To describe the scattering of incident light over space and wavevector, we introduce the `average light field transmission function', $P_W$. This function maps the incident light field $W_a(\xa,\ka)$ of the input field to the ensemble averaged transmitted light field $\langle W_b(\xb,\kb)\rangle$ at the target plane:
	\begin{equation}\label{eq:PW-introduction}
	\begin{split}
	\langle W_b(\xb,\kb)\rangle= \\
	\frac{1}{(2\pi)^2} \iint P_W(\xb, \kb; \xa, \ka) W_a(\xa, \ka) d^2\xa d^2\ka.
	\end{split}
	\end{equation}
	\Eqref{eq:PW-introduction} is the phase space equivalent of \eqref{eq:TM-definition}, now considered with an ensemble average. It is well-known that as a transformation that connects two WDFs, $P_W$ takes the form of a double-Wigner distribution~\cite{Bastiaans:79}:
	\begin{equation}\label{eq:PW-definition}
	\begin{split}
	P_W(\xb,\kb;\xa,\ka) \equiv \\
	\iint \left\langle  T(\xb^+, \xa^+) T^*(\xb^-, \xa^-) \right\rangle e^{
		i(\Dxa\cdot\ka - \Dxb\cdot\kb)} d^2\Dxa d^2\Dxb. 
	\end{split}
	\end{equation}
	Although $P_W$ is a function of 4 variables, it only depends upon the scattering system's 2-variable transmission matrix, $T$, and obeys the same properties as a WDF (e.g., realness). 
	
	In \eqref{eq:PW-definition}, we recognize a Fourier transform from $\Dxb$ to $\kb$ and an inverse Fourier transform from $\Dxa$ to $\ka$. Performing the reversed transforms on both sides of \eqref{eq:PW-definition} yields, 
	\begin{equation}
	\begin{split}
	\left\langle T(\xb^+, \xa^+) T^*(\xb^-, \xa^-) \right\rangle = \\
	\frac{1}{(2\pi)^4} \iint P_W(\xb,\kb;\xa,\ka) e^{
		i(- \Dxa\cdot\ka + \Dxb\cdot\kb)} d^2\ka d^2\kb, 
	\end{split}
	\end{equation}
	which can be inserted into \eqref{eq:CW-definition} to arrive at our central result:
	\begin{widetext}
		\begin{equation}\label{eq:PW-Cw}
		C_W(\Dxb, \Dkb; \Dxa, \Dka)=
		\frac{1}{(2\pi)^4} \iiiint P_W(\xb,\kb;\xa,\ka) e^{i(-\Dxa\cdot\ka + \Dxb\cdot\kb + \Dka\cdot\xa -\Dkb\cdot\xb)} d^2\ka d^2\xa d^2\kb d^2\xb.
		\end{equation}
	\end{widetext}
	
	In short, $P_W$ and $C_W$ are connected through two forward and two inverse 2D Fourier transforms. The correlation function in \eqref{eq:PW-Cw} is the formulation of our new generalized memory effect, generalizing the well-known `tilt' memory effect into a full class of interrelated shift and tilt correlations. At this point we want to emphasize that \eqref{eq:PW-Cw} is still valid for any scattering medium or geometry. Also, these spatio-angular correlations are an intrinsic property of the scattering medium and will thus be present regardless of the form of the input field.
	
	\section{Approximate solution for forward scattering \label{sect:Fokker-Planck}}
	We are now left with the challenge of calculating $P_W$. Although we could use the radiative transfer equation in the most general case, we prefer an approximate solution that gives a simple analytical form for $P_W$ under the condition that light is mainly scattered in the forward direction. Furthermore, we assume paraxial light propagation, as the WDF's joint description of light across $\xx$ and $\kk$ is very closely related to the light field description of rays~\cite{Zhang:2009}. The idea is to approximate the specimen as a series of thin phase plates. Each phase plate slightly changes the propagation direction of the rays while maintaining their position. Between the phase plates, the light propagates along straight rays and maintains its directionality. In the limit of infinitely small steps, this picture translates to a Fokker-Planck equation with the following solution~\cite{liu:2015} (full derivation in the Supplementary Information A):
	\begin{equation}\label{eq:PW-scattering-slab}
	P_W^{FP}(\hat{\xx}, \hat{\kk})  = \frac{12\ell_{tr}^2}{k_0^2 L^4}\mathrm{exp}\left(
	-\frac{6\ell_{tr}}{L}
	\left[
	\frac{|\hat{\xx}|^2}{L^2}
	-\frac{\hat{\kk} \cdot \hat{\xx}}{k_0L}
	+\frac{|\hat{\kk}|^2}{3k_0^2}
	\right] \right).
	\end{equation}
	Here, $\hat{\kk}\equiv\kb-\ka$, $\hat{\xx}\equiv\xb-\xa- L \ka/ k_0$, $k_0$ is the wavenumber, $\ell_{tr}$ is the transport mean free path, and $L$ is the separation between the input and target plane (that is, the target plane depth). 
	
	Interestingly, $P_W^{FP}$ is only a function of two variables. The reduction of $\xb$ and $\xa$ to a single difference coordinate $\hat{\xx}$ is possible because the Fokker-Planck equation is shift invariant. A similar simplification was used in the original derivation of the `tilt' optical memory effect~\cite{Feng:1988}. By assuming the average scattered intensity envelope only depended upon relative position, the resulting memory effect correlation reduced to a function of only one tilt variable, which is now a commonly applied simplification in many experiments~\cite{Katz:2012, Bertolotti:2012, Katz:2014, Yang:2014}. Note that in this paraxial model the target intensity distribution is additionally offset by $L\ka/k_0$, which is exactly what is expected from pure ballistic propagation through a transparent medium of thickness $L$. Moreover, the Fokker-Planck model is also invariant to a tilt in the incident wave. This symmetry allows for the reduction of coordinates $\kb$ and $\ka$ to $\hat{\kk}$. Of course, this approximation neglects the fact that rays at a high incident angle propagate a larger distance inside the sample.
	
	We can now find an expression for the generalized memory effect in a forward scattering material by inserting \eqref{eq:PW-scattering-slab} into \eqref{eq:PW-Cw} (see Supplementary Information A for the details), arriving at
	\begin{equation}\label{eq:invariantCW}
	\begin{split}
	C_W(\Dxb, \Dkb; \Dxa, \Dka)= (2\pi)^2 \\
	C_W^{FP}(\Dxb, \Dkb) \delta(\Dka-\Dkb) \delta(\Dxb - \Dxa - \Dka L/k_0).
	\end{split}
	\end{equation}
	Here the 2D-correlation function $C_W^{FP}$ is given by 
	\begin{equation}\label{eq:CW-scattering-slab}
	\begin{split}
	C_W^{FP}(\Dxb, \Dkb) =\\
	\exp\left(
	-\frac{L^3k_0^2}{2\ell_{tr}}
	\left[
	\frac{|\Dkb|^2}{3k_0^2}
	-\frac{\Dkb\cdot\Dxb}{k_0L}
	+\frac{|\Dxb|^2}{L^2}
	\right]\right).
	\end{split}
	\end{equation}
	
	The two delta functions in \eqref{eq:invariantCW} are a direct result of the shift and tilt invariance of the Fokker-Planck model. As we show in Supplementary Information B, in an actual experiment the delta functions will be replaced by the ambiguity function of the incident field, which is ideally a well-behaved sharp function.
	
	As is clear from the cross-term in \eqref{eq:CW-scattering-slab}, shift and tilt correlations along the target plane are not independent but show a combined effect. For many applications, it is useful to shift the field in the target plane as far as possible while maintaining a maximum amount of correlation. To achieve this, one should simultaneously tilt and shift the incident field to maximize $C_W^{FP}$ for a desired shift distance $\Dxb$, while using $\Dkb$ as a free parameter. From \eqref{eq:CW-scattering-slab}, we find that the optimal scan range is achieved when 
	\begin{equation}\label{eq:optimal_dkb}
	\Dkb^{opt} = \frac{3k_0\Dxb}{2L}.
	\end{equation}
	We discuss the implications of this finding to adaptive optics in Section~\ref{sect:isoplanatism}.
	
	\section{Experimental validation \label{sect:experiments}}
	We now measure $C_W$ and $P_W$ for a forward scattering medium in two separate experiments. These experiments will verify the $P_W$-$C_W$ relation predicted in \eqref{eq:PW-Cw} and the Fokker-Planck model in \eqref{eq:PW-scattering-slab}. For simplicity, we consider $C_W$ and $P_W$ only along the horizontal dimension, with $\text{x}_b$ and $\text{k}_b$ as the horizontal components of the position and the wavevector at the sample back surface, respectively. We created two samples of $3 \mu$m-diameter silica microspheres immersed in agarose gel with thicknesses of $L=258 \pm 3 \mu$m and $520 \pm 5$ $\mu$m. Using Mie theory, we find an anisotropy factor of $g=0.980 \pm 0.007$  and a scattering mean free path of $l_{sc}=0.296 \pm 0.016$ mm at a wavelength of 632.8 nm, yielding a transport mean free path of $\ell_{tr}=14.8 \pm 5.2$ mm.

	In the experimental setup, light from a 632.8 nm HeNe laser is expanded and split into a reference path and the sample path. The light is focused onto the scattering sample using an objective (100x/0.8), and we image the sample backsurface (target plane) with a CCD camera through a second identical objective and tube lens (f = 150mm). The phase of light transmitted through the sample is determined by means of off-axis holography. The sample is placed on a translation stage for lateral movement. Additionally in the experiments, either a pinhole or diffuser is placed in front of the first microscope objective. The experimental setup is illustrated in the Supplementary Figure.
	
	\subsection{Measurements of the average light field transmission function}
	From \eqref{eq:PW-introduction}, we know that a finite input beam $E_a$ will result in an average WDF at the target plane, $\left<W_b\right>$, that is a convolution between the desired average light field transmission function $P_W$ and the WDF of the input field, $W_a$. We are thus able to determine $P_W$ by using a pencil beam for sample illumination (size = 2.0 $\mu$m) at $\text{x}_a =0$ and $\text{k}_a=0$, which approximately forms the input WDF $W_a(\text{x}_a,\text{k}_a) \propto \delta(\text{x}_a) \delta(\text{k}_a)$. $P_W$ is then found by measuring the average scattered light at the target plane across both space and angle. The pencil beam is formed by placing a 500 $\mu$m-wide pinhole close to the back-aperture of the first objective. After measuring the scattered field, we numerically calculate the WDF $W_b(\text{x}_b,\text{k}_b)$, using \eqref{eq:Wigner-definition}, to find the scattered intensity as a function of both $\text{x}_b$ and $\text{k}_b$. We average the WDF over 300 measurements, translating the sample over a distance of 10 $\mu$m in between each measurement for statistical variation. To facilitate comparison of our measurements with the Fokker-Planck model, which does not include ballistic light, we chose to remove the contribution of ballistic light by subtracting the average transmitted field from every measured field before calculating its WDF. Following prior work, it is possible to modify our model to also include ballistic light at the expense of some added complexity ~\cite{Wax1998, cheng2000propagation}. 
	
\begin{figure}
		\includegraphics[width=1\columnwidth]{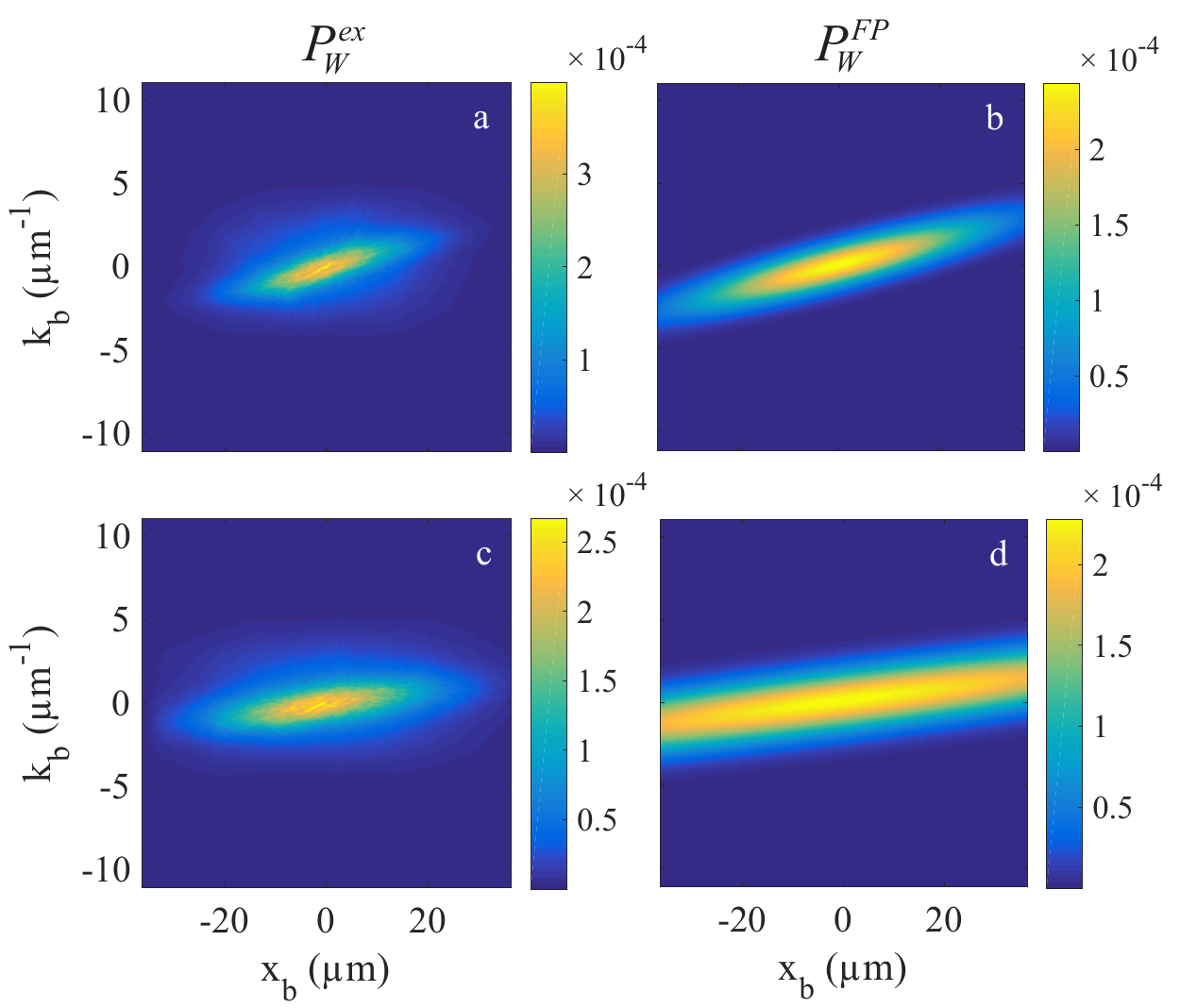}
		\caption{Results of the average light field transmission function ($P_W$) experiment. We compare our measurements, $P_W^{ex}$, to the Fokker-Planck model prediction ($P_W^{FP}$, from \eqref{eq:PW-scattering-slab}) for samples with (\textbf{a-b}) $L = 258$ $\mu$m and (\textbf{c-d}) $L = 520$ $\mu$m. Colorbar indicates the normalized transmitted intensity as function of $\text{x}_b$ and $\text{k}_b$.}
		\label{fig:PW-results}
	\end{figure}

	In Fig.~\ref{fig:PW-results}, we compare our measured average light field transmission functions, $P_W^{ex}$, to those computed with the Fokker-Plank model ($P_W^{FP}$) in \eqref{eq:PW-scattering-slab}. The effect of the cross-term between $\text{x}_b$ and $\text{k}_b$, as predicted in $P_W^{FP}$, is clearly visible in our measurements from the fact that the distributions are sheared. This shear implies that the light at the edges of the diffuse spot (large $\text{x}_b$) continues to propagate, on average, in a radially outward manner (large $\text{k}_b$) after scattering. The $P_W^{ex}$ measurements are less spread out than the Fokker-Planck model, which might be a result of the limited optical sensitivity at the edges of the objective lenses. Both a large field of view and a large numerical aperture of the microscope objective are required to measure the full extent of the average light field transmission function.
	
	\begin{figure*}[t]
		\includegraphics[width=1\textwidth]{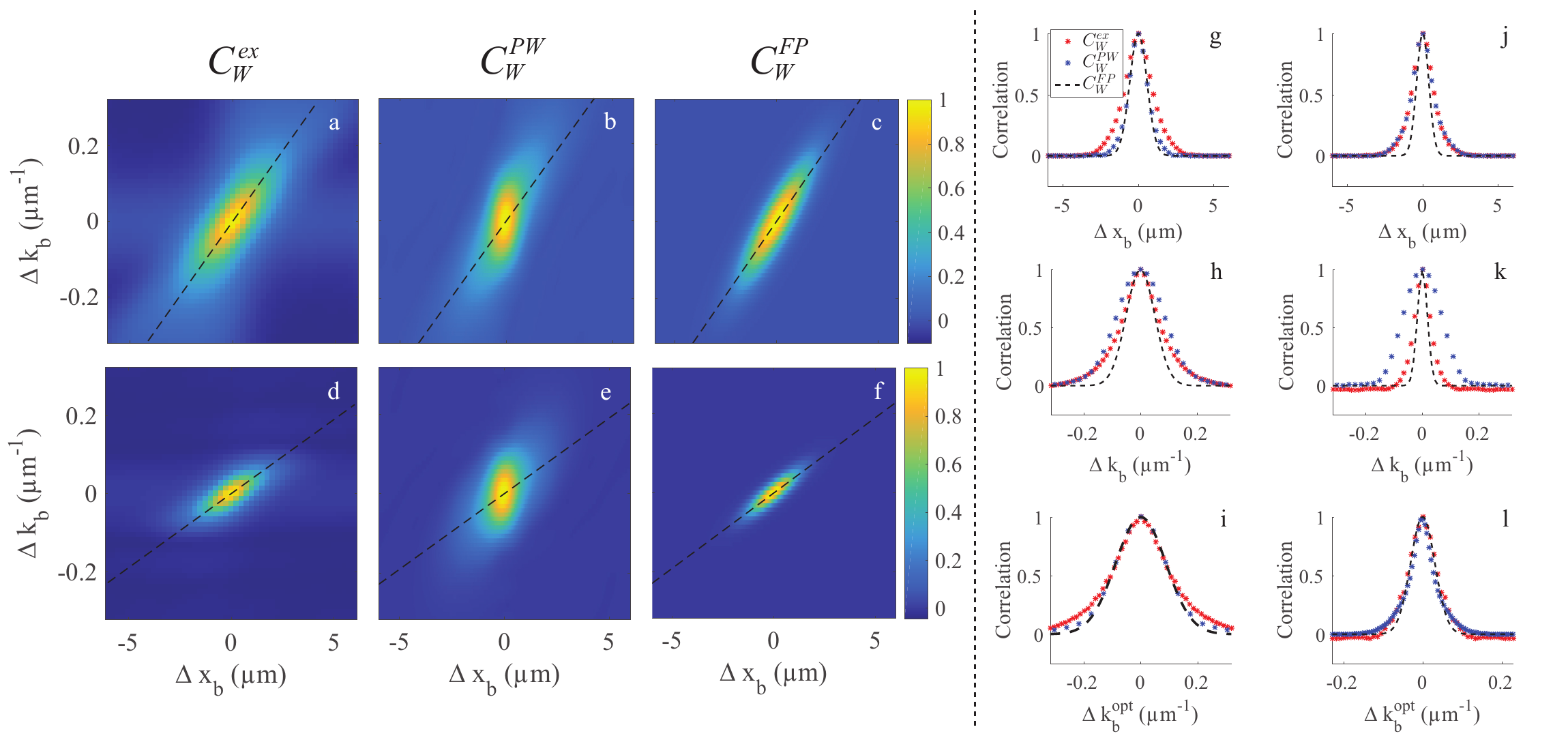}
		\caption{Results of the generalized memory effect correlation ($C_W$) experiments. Measurements of $C_W^{ex}$ are compared to $C_W^{PW}$ and the Fokker-Planck correlation model $C_W^{FP}$ from $\eqref{eq:CW-scattering-slab}$ for samples with $L = 258$ $\mu$m (\textbf{a-c}) and $L = 520$ $\mu$m (\textbf{d-f}). Dashed lines indicate the optimal scanning condition in $\eqref{eq:optimal_dkb}$. (\textbf{g-i}), Cross sections of the 2D correlation functions in (\textbf{a-c}), evaluated (\textbf{g}) along the horizontal axis at $\Dkb=0$, (\textbf{h}) along vertical axis at $\Dxb=0$, and (\textbf{i}) along the optimal scan line for the 258~$\mu$m-thick sample. (\textbf{j-l}), Same cross sections for the 520~$\mu$m-thick sample. The black dashed line denotes the $C_W^{FP}$ model, while red and blue stars denote the measured $C_W^{ex}$ and $C_W^{PW}$, respectively.}
		\label{fig:CWresults}
	\end{figure*}
	
	\subsection{Measurements of the generalized correlation function}
	Next, we experimentally measure the generalized correlation function $C_W^{ex}$. We will directly compare these measurements to our correlation model for $C_W^{FP}$ in \eqref{eq:CW-scattering-slab}. We will also compare these measurements to the 2D Fourier transform of the average light field transmission functions in Fig.~\ref{fig:PW-results}, which will verify our main result in \eqref{eq:PW-Cw}. For the correlation measurements, we replace the pinhole in front of the first objective with a diffuser that forms a randomized input field (average speckle size = 400 nm) at the sample surface. We use a diffuser to minimize correlations within the input field, which manifest themselves as a convolution in our measurement of $C_W^{ex}$ (see Supplementary Information B). We tilt the random input field by translating the diffuser at the objective back aperture. We record a total of 625 scattered fields by illuminating the sample at 25 unique spatial locations and under 25 unique angles of incidence. From this $25^2$ data cube, we compute $C_W^{ex}$ by taking the ensemble average of all the absolute values of the correlation coefficients between two fields separated by the same amount of shift $\Delta \text{x}_b$ and tilt $\Delta \text{k}_b$. Finally, we normalize $C_W^{ex}$ after subtracting the correlation value at the maximum shift, corresponding with the correlations in the ballistic light.
	
	Fig.~\ref{fig:CWresults} presents the results of our $C_W$ experiments. For the two different sample thicknesses of $L = 258$ $\mu$m and $L = 520$ $\mu$m, the measured 2D correlation functions in (a,d) are compared to $C_W^{PW}$, the Fourier transform of $P_W^{ex}$ in (b, e), and the Fokker-Planck model in (c,f). The dashed line indicates the optimal tilt/scan condition as predicted in $\eqref{eq:optimal_dkb}$. We also show cross sections through $C_W^{ex}$ (red stars), $C_W^{PW}$ (blue stars) and $C_W^{FP}$ (dashed black) on the right. These are given along the $\Delta \text{k}_b=0$ line (g,j), the $\Delta \text{x}_b=0$ line (h,k), and the optimal target plane scan condition (i,l). These last two plots demonstrate how jointly considered tilts and shifts can increase correlations along $\Delta \text{x}_b$ to increase the scan range (i.e., isoplanatic patch) at the target plane. Our correlation measurements in Fig.~\ref{fig:CWresults} closely match the Fokker-Planck model. The cross sections for $C_W^{ex}$ and $C_W^{PW}$ are also in good agreement, except for Fig.~\ref{fig:CWresults}k. This data corresponds to the thickest sample, for which the limited field of view of the objective lens prevented us from fully measuring $P_W^{ex}$ along $\text{x}_b$. The truncation of $P_W^{ex}$, already observed in Fig.~\ref{fig:PW-results}c, results in a broadening of its Fourier transform $C_W^{PW}$. 
	
	\section{Maximizing the isoplanatic patch \label{sect:isoplanatism}}
\begin{figure}
		\includegraphics[width=\columnwidth]{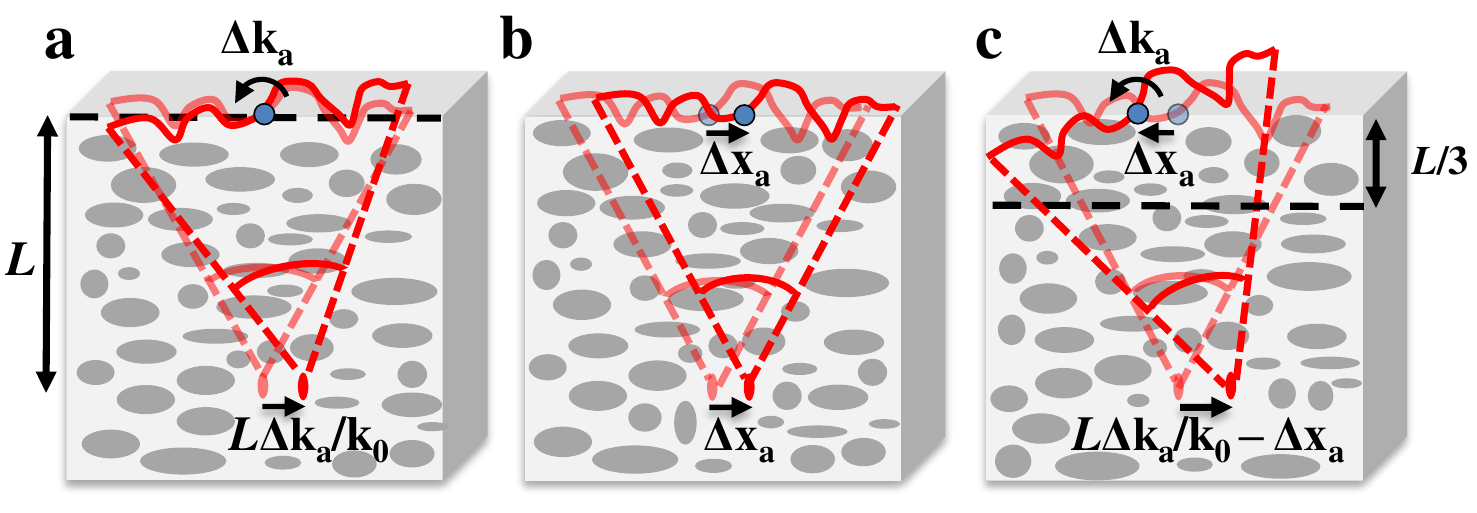}
		\caption{AO focus scanning/imaging inside a scattering medium uses different memory effects. (\textbf{a}) The `tilt' effect arises with the AO tilt plane  (dashed line) conjugated to the input surface, (\textbf{b}) the `shift' effect arises with the AO tilt plane at infinity, and (\textbf{c}), the optimal joint tilt/shift scheme requires the AO tilt plane to be at a distance $L/3$ inside the sample.}
		\label{fig:scanning}
	\end{figure}

	Having experimentally confirmed the validity of the generalized memory effect in forward scattering media, we now examine its application to adaptive optics (AO) systems. AO allows light to be focused inside scattering media by correcting for the induced wavefront distortions by means of a deformable mirror or spatial light modulator. Generally, AO systems are limited to a single plane of wavefront correction which can be tilted to scan the focus. However, a single correction plane cannot correct for a full scattering volume and therefore the focus scan range is limited to a small area termed the isoplanatic patch. A central question in AO is where to conjugate the correction plane to maximize the isoplanatic patch~\cite{Mertz2015}. In the case of conjugate AO, this correction plane is located at the sample's top (input) surface, whereas in pupil AO it is effectively located at an infinite distance from the sample. In Fig.~\ref{fig:scanning}, we diagram how conjugate AO and pupil AO are analogous to experiments that measure the `tilt' (Fig.~\ref{fig:scanning}a) and `shift' memory effects (Fig.~\ref{fig:scanning}b), respectively. Put in terms of our model, maximizing the isoplanatic patch equates to finding an optimal combination of $\Dxa$ and $\Dka$ to maximize $C_W^{FP}$ at a shifted target position $\Dxb$.  
	
	In $\eqref{eq:CW-scattering-slab}$, we considered $C_W^{FP}$ only in terms of translation and rotation of the target field $\Dxb$ and $\Dkb$. However, $C_W^{FP}$ can also be expressed as a function of the translation and rotation of the incident field $\Dxa$ and $\Dka$. Using the delta function relations from \eqref{eq:invariantCW}, we can substitute $\Dxb = \Dxa + L\Dka/k_0$ and $\Dkb = \Dka$ in $\eqref{eq:optimal_dkb}$ to find the optimal tilt/shift combination at the input plane for a given amount of target shift $\Dxb$:
	\begin{equation}\label{eq:optimal_dka}
	\Dka^{opt} = \frac{3k_0\Dxb}{2L} \text{ and } \Dxa^{opt} = -\Dxb/2.
	\end{equation}
	Hence, to scan a focus by a distance of $\Dxb$, the optimal strategy is to shift the incident field by $\Dxb/2$ \emph{in the opposite direction} and then tilt until the desired $\Dxb$ is reached. In other words, the optimal scanning mechanism is achieved by conjugating the AO correction plane inside the scattering sample at a depth of $L/3$, which geometrically corresponds to the ideal tilt/shift rotation plane. We illustrate this mechanism for optimized focus scanning in Fig.~\ref{fig:scanning}c. 
	
	In Table~\ref{tab:scanning}, we compare the isoplanatism provided by the three different memory effects. Here, we define isoplanatism as the maximum allowed scan distance $\xb$ at the target plane that maintains $1/e$ correlation, which we solve for with appropriately defined input variables in $\eqref{eq:CW-scattering-slab}$. Interestingly, all memory effects decrease at the same rate of $\sqrt[]{\ell_{tr}/k_0^2L}$ as the target plane is placed deeper inside the medium. However, the scan range of the tilt/tilt memory effect is always a factor of $\sqrt[]{3}$ larger than the shift/shift memory effect. By exploiting the optimal tilt/shift combination, as given in $\eqref{eq:optimal_dka}$, the scanning improvement is increased to a factor of 2. 
	
	\begin{table}[b]
		\centering
		\begin{tabular}{ | l | l | c | c | }
			\hline
			Memory effect 	& Adaptive Optics 	& Tilt plane	& Scan range \\
			\hline
			Shift			& Pupil				& $-\infty$		& $\sqrt[]{2\ell_{tr}/k_0^2L}$ \\
			Tilt			& Surface Conjugate	& $0$			& $\sqrt[]{6\ell_{tr}/k_0^2L}$ \\ 
			Generalized		& Sample Conjugate	& $L/3$			& $\sqrt[]{8\ell_{tr}/k_0^2L}$ \\
			\hline
		\end{tabular}
		\caption{Comparison of the performance of the three different memory effects in terms of adaptive optics scan range.}
		\label{tab:scanning}
	\end{table}
	
	\section{Discussion \label{sect:Disc}}
	The optical memory effect as reported in Ref.~\cite{Feng:1988} has paved the way for several new imaging techniques that can `see through' thin scattering layers~\cite{Katz:2012, Bertolotti:2012, Katz:2014, Yang:2014}. These techniques require the object of interest to be positioned at a distance behind the thin layer and are thus not immediately applicable to situations in biomedical imaging where the object of interest is embedded in scattering tissue. The anisotropic memory effect~\cite{Judkewitz:2015} showed that translational spatial correlations also exists within the scattering sample, which have been demonstrated to be suitable for imaging or focus scanning inside biological tissue~\cite{Papadopoulos2016,Park2015}. Our new generalized memory effect model offers a new theoretical framework, combining the two known memory effects into a single complete model and offering a means to increase the imaging/scanning area. 
	
	Using the Fokker-Planck light propagation model, we have shown that the `tilt' memory effect is not only present behind, but also \emph{inside} scattering layers, proving to be a factor $\sqrt[]{3}$ more effective as a scanning technique than the `shift' memory effect alone. This finding supports the field-of-view (FOV) advantage of conjugate AO over pupil AO discussed in Ref.~\cite{Mertz2015}, as the `tilt' and `shift' memory effects are utilized by conjugate AO and pupil AO, respectively. Our optimal joint tilt/shift scheme, which corresponds to an optimal AO conjugation tilt plane, maximizes the corrected FOV beyond what is predicted independently by only the `tilt' memory effect and only the `shift' memory effect. The scan ranges given in Table~\ref{tab:scanning} are strictly the correlations that result from the transmission matrix. Correlations in the input beam further extend the scan range of the different memory effects (see Supplementary Information B). The Fokker-Planck model assumes continuous scattering throughout the sample and neglects back-scattered light, which might not hold in strong isotropic scattering media. However, in this case the Fourier relation between $P_W$ and $C_W$ is still valid. The generalized correlation function is the phase-space equivalent of the $C_1$ intensity correlations introduced by Feng et al.~\cite{Feng:1988}. We envisage that there also exist phase-space equivalents of $C_2$ and $C_3$ correlations. These higher-order correlations have recently been exploited to focus light through scattering media with an unexpectedly high efficiency, and investigating their phase-space equivalents may prove equally useful \cite{Hsu:2017}.
	
	Concluding, our new generalized memory effect model generalizes the known optical memory effects without making any assumptions about the geometry or scattering properties of the scattering system. The predicted Fourier relation between the average light field transmission function and generalized correlation function has been experimentally verified in forward scattering media. Furthermore, we found that the simple Fokker-Planck model for light propagation is surprisingly accurate in describing the full set of first-order spatial correlations inside forward scattering media. Our new generalized memory effect model predicts the maximum distance that a scattered field can be scanned while remaining correlated to its unshifted form. In other words, the generalized memory effect provides the optimal scan mechanism for deep-tissue focusing techniques.

\clearpage

\onecolumngrid

\section*{Supplementary Information A: Fokker-Planck model \label{app:Fokker-Planck}}
In this section, we derive the Fokker-Planck equation used to model the generalized memory effect, and show how it leads to the simple two-dimensional correlation function over space and angle in \eqref{eq:CW-scattering-slab}. We use a paraxial model to describe light propagation inside a forward scattering medium. We assume that light rays propagate through the medium at an angle $\arcsin(\kk / k_0)\approx \kk / k_0$, with $\kk \equiv (k_x, k_y)$ the perpendicular component of the wave vector.
The scattering sample is modeled as a stack of scattering phase plates~\cite{liu:2015}, where each plate slightly changes the direction of the beam (i.e. $\kk$), while preserving the position. Between the phase plates, the rays propagate along straight trajectories. In the limit where the number of phase plates per unit length becomes infinite, this step-scatter model reduces to the following Fokker-Planck type partial differential equation for the radiance:
\begin{equation}\tag{A1}
\begin{split}
\frac{\partial I(\xx, \kk, z)}{\partial z} =
-\frac{k_x}{k_0}\frac{\partial I(\xx, \kk, z)}{\partial x}
-\frac{k_y}{k_0}\frac{\partial I(\xx, \kk, z)}{\partial y}
+\beta k_0^2\left(\frac{\partial^2 I(\xx, \kk, z)}{\partial k_x^2}
+\frac{\partial^2 I(\xx, \kk, z)}{\partial k_y^2}
\right),
\label{eq:Fokker-Planck}
\end{split}
\end{equation}
where $\xx \equiv (x, y)$ and $z$ is the optical axis. Here, the first two terms correspond to propagation of light along straight trajectories, while the last two terms describe scattering. We will find an expression for the scattering strength $\beta$ later.

The model in \eqref{eq:Fokker-Planck} neglects all backscattered light. Furthermore, the use of a Fokker-Planck equation (also known as a Kolmogorov forward equation) implies a continuous diffusion process, i.e., even after a very small propagation distance all light has been scattered, albeit at a very small angle. Physically, this model corresponds to having the limit of the scattering length approach zero, while keeping fixed the transport length. The important implication of this assumption is that there is no ballistic light in the Fokker-Planck model.

Under the initial condition $I(\xx, \kk, z=0)= (2\pi)^2 \delta(\kk-\ka)\delta(\xx-\xa)$, we find the following solution at $z=L$ using the method outlined in Ref.~\cite{Tanski:2004}
\begin{equation}\label{eq:PW-sol-1}\tag{A2}
I(\hat{\xx},\hat{\kk}, z=L)=
\frac{3}{ k_0^2 \beta^2 L^4}
\exp{\left(
	-\frac{3}{\beta L}
	\left[
	\frac{|\hat{\xx}|^2}{L^2}
	-\frac{\hat{\xx}\cdot\hat{\kk}}{L k_0}
	+\frac{|\hat{\kk}|^2}{3k_0^2}
	\right]
	\right)}.
\end{equation}
Here, we introduced the variables $\hat{\kk}\equiv\kk-\ka$ and $\hat{\xx}\equiv\xx-\xa- L \ka/ k_0$. It can be easily verified through substitution that \eqref{eq:PW-sol-1} is a (normalized) solution to \eqref{eq:Fokker-Planck}. The difference coordinate variables $\hat{\kk}$ and $\hat{\xx}$ now define the two dimensional light field transmission function $P_W(\hat{\xx},\hat{\kk},)\equiv I(\hat{\xx},\hat{\kk}, L)$ that we introduce in \eqref{eq:PW-scattering-slab}.

It is worth noting that $I(\hat{\xx},\hat{\kk}, L)$ describes a joint Gaussian distribution along $\hat{\xx}$ and $\hat{\kk}$ with covariance matrix,
\begin{equation}\tag{A3}
\Sigma=\beta L\begin{bmatrix}
\frac{2 L^2}{3} & 0 & k_0 L & 0\\
0 & \frac{2 L^2}{3} & 0 & k_0 L\\
k_0 L & 0 & 2 k_0^2 & 0\\
0 & k_0 L & 0& 2 k_0^2.
\end{bmatrix}
\end{equation}
At this point, we still need to determine the scattering strength $\beta$. In order to do so, we compare our results to a discrete scattering model, which assumes that light propagates an average distance of $l_{sc}$ between each scattering event (i.e., $l_{sc}$ is the medium's scattering coefficient).

For this comparison, we first use~\eqref{eq:PW-sol-1} to calculate the average angular spread of light, $\langle|\hat{\kk}|^2\rangle$, as a function of depth:
\begin{equation}\label{eq:kb-var-beta}\tag{A4}
\langle|\hat{\kk}|^2\rangle \equiv \frac{1}{(2\pi)^2} \iint |\hat{\kk}|^2 I(\hat{\xx},\hat{\kk},L) d\hat{\xx} d\hat{\kk} = 4 k_0^2 L \beta.
\end{equation}
Here, we see that the variance of the angular distribution grows linearly with propagation distance $L$, consistent with a diffusion process in the $\kb$ coordinate system at the target plane.

Second, we calculate the expected angular spread as a function of depth for the discrete scattering model. After propagating a short distance $L\ll l_{sc}$, a fraction of $1-\exp {(L/l_{sc})}$ of the light is scattered. The angular spread of the scattered light is usually quantified with the anisotropy factor $g\equiv \langle\cos{\theta}\rangle$, with $\theta$ the scattering angle. Of course, the non-scattered light will still maintain its original direction. When $L$ is small, we can write $g=\langle\cos\theta\rangle\approx 1-\langle|\kb|^2\rangle/(2k_0^2)$ and we find that
\begin{equation}\label{eq:kb-var-g}\tag{A5}
\langle |\hat{\kk}|^2 \rangle = 2k_0^2(1-e^{-L/l_{sc}})(1-g)\approx 2k_0^2 L \frac{1-g}{l_{sc}}.
\end{equation}
In the limit of $L\rightarrow 0$, we can combine \eqref{eq:kb-var-beta} and \eqref{eq:kb-var-g} to find
\begin{equation}\tag{A6}
\beta = \frac{1-g}{2 l_{sc}} = \frac{1}{2 l_{tr}},
\end{equation}
with $l_{tr}\equiv l_{sc}/(1-g)$ denoting the transport mean free path. Substituting this result into \eqref{eq:PW-sol-1} finally gives us equation (8) in the main text.

To find $C_W^{FP}$, we need to evaluate \eqref{eq:PW-Cw} using the solution for $P_W = P_W^{FP}$ in \eqref{eq:PW-scattering-slab}. Since \eqref{eq:PW-scattering-slab} only depends on the difference coordinates $\hat{\xx}$ and $\hat{\kk}$, we first rewrite the integrals over $\xb$ and $\kb$ to integrals over $\hat{\xx}$ and $\hat{\kk}$ by applying the following changes of variables: $\xb\rightarrow \hat{\xx}+\xa+L\ka/k_0$ and $\kb\rightarrow \hat{\kk}+\ka$. Doing so leads to,
\begin{equation}\tag{A7}
\begin{split}
C_W(\Dxb, \Dkb; \Dxa, \Dka)= \\
\frac{1}{(2\pi)^4} \iiiint P_W^{FP}(\hat{\xx}, \hat{\kk}) e^{i(-\Dxa\cdot\ka + \Dxb\cdot(\hat{\kk}+\ka) + \Dka\cdot\xa -\Dkb\cdot(\hat{\xx}+\xa+L\ka/k_0))} d^2\hat{\kk} d^2\hat{\xx} d^2\ka d^2\xa = \\
\frac{1}{(2\pi)^4} \iint P_W^{FP}(\hat{\xx}, \hat{\kk}) e^{i(\Dxb\cdot\hat{\kk}-\Dkb\cdot\hat{\xx})} d^2\hat{\kk} d^2\hat{\xx}
\int e^{i(\Dxb-\Dxa-\Dkb L/k_0) \cdot\ka} d^2\ka \int e^{i(\Dka-\Dkb)\cdot\xa} d^2\xa.\label{eq:CW4-to-2}
\end{split}
\end{equation}
The integrals over $\ka$ and $\xa$ reduce to the delta function product $(2\pi)^4 \delta(\Dxb-\Dxa-\Dkb L /k_0) \delta(\Dka-\Dkb)$, or, equivalently $(2\pi)^4 \delta(\Dxb-\Dxa-\Dka L /k_0) \delta(\Dka-\Dkb)$. We can now write the original four dimensional correlation function as,
\begin{equation}\tag{A8}
C_W(\Dxb, \Dkb; \Dxa, \Dka) = (2\pi)^2\delta(\Dxb-\Dxa-\Dka L /k_0) \delta(\Dkb-\Dka) C_W^{FP}(\Dxb, \Dkb),
\end{equation}
with
\begin{equation}\label{eq:CW2D}\tag{A9}
C_W^{FP}(\Dxb, \Dkb) \equiv \frac{1}{(2\pi)^2} \iint P_W^{FP}(\hat{\xx}, \hat{\kk}) e^{i(\Dxb\cdot\hat{\kk}-\Dkb\cdot\hat{\xx})} d^2\hat{\kk} d^2\hat{\xx}.
\end{equation}
Here, $C_W^{FP}$ is a generalized correlation function that is the forward and inverse 2D Fourier transforms along $\hat{\xx}$ and $\hat{\kk}$ of the associated average light field transmission function, $P_W^{FP}(\hat{\xx}, \hat{\kk})$, for the shift and tilt invariant medium. We use $P_W=P_W^{FP}$ for the particular case of comparing our results to the Fokker-Planck model, described in \eqref{eq:CW-scattering-slab}. However, \eqref{eq:CW2D} holds for any shift/tilt invariant light field transmission function.
In the case of the Fokker-Planck model, the total transmission is unity, and the correlation coefficient in~\eqref{eq:CW2D} is normalized. Otherwise, we may choose to normalize the correlation coefficient by dividing by $\frac{1}{(2\pi)^2} \iint P_W(\hat{\xx}, \hat{\kk}) d^2\hat{\kk} d^2\hat{\xx}$, which is the total transmissivity of the sample.

\section*{Supplementary Information B: Correlation measurement \label{app:CWmeasurement}}
In this section, we present additional mathematical details about our second experiment, where we directly measured the generalized memory effect correlation function. We performed our correlation measurements by first measuring many scattered field responses at the target plane for an input speckle field that we both shift and tilt via translation stages. Here, the sample is physically shifted by a distance $\Dx$ and the incident wavefront is tilted by $\Dk$ by translating the diffuser in front of the microscope objective.

Hence, a single shifted/tilted measurement will take the form,
\begin{equation}\label{eq:measurement}\tag{B1}
E^{ex}_{\Dx, \Dk}(\xb) = \int T(\xb-\Dx, \xx_{a}-\Dx) E(\xx_{a}) 
e^{i\Dk\cdot\xx_{a}} d^2 \xa.
\end{equation}
We then perform a pairwise analysis of the measured fields. For each pair of fields, we have $\Dxa = \Dx_1-\Dx_2$ and $\Dka = \Dk_1-\Dk_2$. For simplicity, and without loss of generality, we now only consider the symmetric case $\Dx_1=-\Dx_2=2 \Dxa$ and $\Dk_1=-\Dk_2=2 \Dka$.

After recording, we digitally shift and tilt the first field by $-\Dxb/2$ and $-\Dkb/2$, respectively, and we shift and tilt the second field by $\Dxb/2$ and $\Dkb/2$. Finally, we determine the overlap of the two fields, effectively calculating
\begin{equation}\label{eq:experimental-CW}\tag{B2}
\begin{split}
C_W^{ex}(\Dxb, \Dkb; \Dxa, \Dka) = 
\int
\left[
\int T(\xb^+, \xx_{a'}+\Dxa/2) E(\xx_{a'}) 
e^{i\Dka \cdot \xx_{a'}/2}
d^2 \xx_{a'}
\right] \times \\
\left[
\int T(\xb^-, \xx_{a''}-\Dxa/2) E(\xx_{a''}) 
e^{-i\Dka \cdot \xx_{a''}/2}
d^2 \xx_{a''}
\right]^*
e^{-i\Dkb \cdot \xb}
d^2 \xb,
\end{split}
\end{equation}
with $\xb^\pm\equiv \xb\pm\Dxb/2$.
At this point, we wish to distinguish the effects of correlations in the transmission matrix and correlations in the incident field. For this purpose, we introduce the ambiguity function~\cite{Zhang:2009}
\begin{equation}\tag{B3}
A(\bkappa, \brho) \equiv \int E(\xa+\brho/2) E^*(\xa-\brho/2) e^{-i\bkappa \cdot \xa} d^2 \xa.
\end{equation}
By applying an inverse Fourier transform, we find
\begin{equation}\tag{B4}
E(\xa+\brho/2) E^*(\xa-\brho/2) = 
\frac{1}{(2\pi)^2}\int A(\bkappa, \brho) e^{i\bkappa \cdot \xa}d^2\bkappa.
\end{equation}
Next, we identify $\xx_{a'}=\xa+\brho/2$ and $\xx_{a''}=\xa-\brho/2$ so that we can substitute this result into \eqref{eq:experimental-CW}
\begin{equation}\tag{B5}
\begin{split}
C_W^{ex}(\Dxb, \Dkb; \Dxa, \Dka) =
\frac{1}{(2\pi)^2}\iiiint
T(\xb^+, \xa+\brho/2+\Dxa/2) \times \\
T^*(\xb^-, \xa-\brho/2-\Dxa/2) 
e^{i(\Dka \cdot \xa -\Dkb \cdot \xb+\bkappa \cdot \xa)}
A(\bkappa,\brho)
d^2\xb
d^2\xa
d^2\bkappa
d^2\brho.
\end{split}
\end{equation}
We can now substitute the definition of $C_W$ in \eqref{eq:CW-definition} to arrive at

\begin{equation}\label{eq:CWconvolution}\tag{B6}
\begin{split}
C_W^{ex}(\Dxb, \Dkb; \Dxa, \Dka) = \frac{1}{(2\pi)^2} \iint C_W(\Dxb, \Dkb; \Dxa+\brho, \Dka+\bkappa) A(\bkappa, \brho) d^2\bkappa d^2\brho.
\end{split}
\end{equation}
In words, the measured correlation function is the actual correlation function of the sample, convolved with the ambiguity function of the incident field mirrored along both $\bkappa$ and $\brho$. 

Let us evaluate \eqref{eq:CWconvolution} for a correlation function for a shift/tilt invariant medium, as we found with the Fokker-Planck model. Substituting \eqref{eq:CW2D} into \eqref{eq:CWconvolution} gives

\begin{equation}\label{eq:CWconvolution2d}\tag{B7}
\begin{split}
C_W^{ex}(\Dxb, \Dkb; \Dxa, \Dka) = \\ \iint C_W^{FP}(\Dxb, \Dkb) \delta(\Dxb-\Dxa-\brho-(\Dka+\bkappa) L /k_0) \delta(\Dkb - \Dka - \bkappa) A(\bkappa, \brho) d^2\brho d^2\bkappa = \\
C_W^{FP}(\Dxb, \Dkb) A(\Dkb-\Dka, \Dxb-\Dxa-\Dkb L / k_0).
\end{split}
\end{equation}

Let us now consider some typical experimental scenarios. First of all, for any incident beam with unit power, the ambiguity function will have a maximum at $A(0,0)=1$. If the incident beam approximates a plane wave, then $A$ will be sharp in the direction of $\bkappa$, and almost constant in the direction of $\brho$. As a result, \eqref{eq:CWconvolution2d} predicts that we can vary $\Dxb$ without observing an effect on the correlation. This result is to be expected since the incident beam is basically translational invariant, shifting it has no effect. This symmetry does not hold for the scattered wave, and thus $C^{ex}_W$ decreases with $\Dxb$ in the same way as $C^{FP}_W$ decreases with $\Dxb$. Finally, the fact that $A$ is sharp in the direction of $\kappa$ makes that tilt/tilt correlations are only observed when $\Dkb\approx\Dka$, as is well-known from experimental observations of the tilt/tilt optical memory effect. 

Second, if we focus the incident light onto the sample with a high numerical aperture, $A$ will be sharp in the direction of $\brho$, and almost constant in the direction of $\bkappa$. In this case, the incident beam is tilt invariant, so we expect $C^{ex}_W$ not to remain constant when the incident beam is tilted (within the numerical aperture).

To summarize, we found that the angular correlations for an incident plane wave are only determined by the sample, whereas for a finite-size beam the correlations in the incident beam give rise to observed correlations over a larger angular range. This result is in line with observations by Li and Genack~\cite{Li:1994}.

By creating a highly randomized input field for our correlation measurements, we were able to create a field with an ambiguity function that is sharp both along $\bkappa$ and $\brho$, thus minimizing the impact of $A$ in this convolution and obtaining an accurate measurement of $C_W$. Additionally, in our experiment, we normalize the measured correlation coefficient and take the average over all pairs of measurement with the same spacing $\Dxa$ and $\Dka$ to average over disorder. 

\clearpage
\section*{Supplementary Figure}

\begin{figure*}[h]
	\includegraphics[width=0.7\columnwidth]{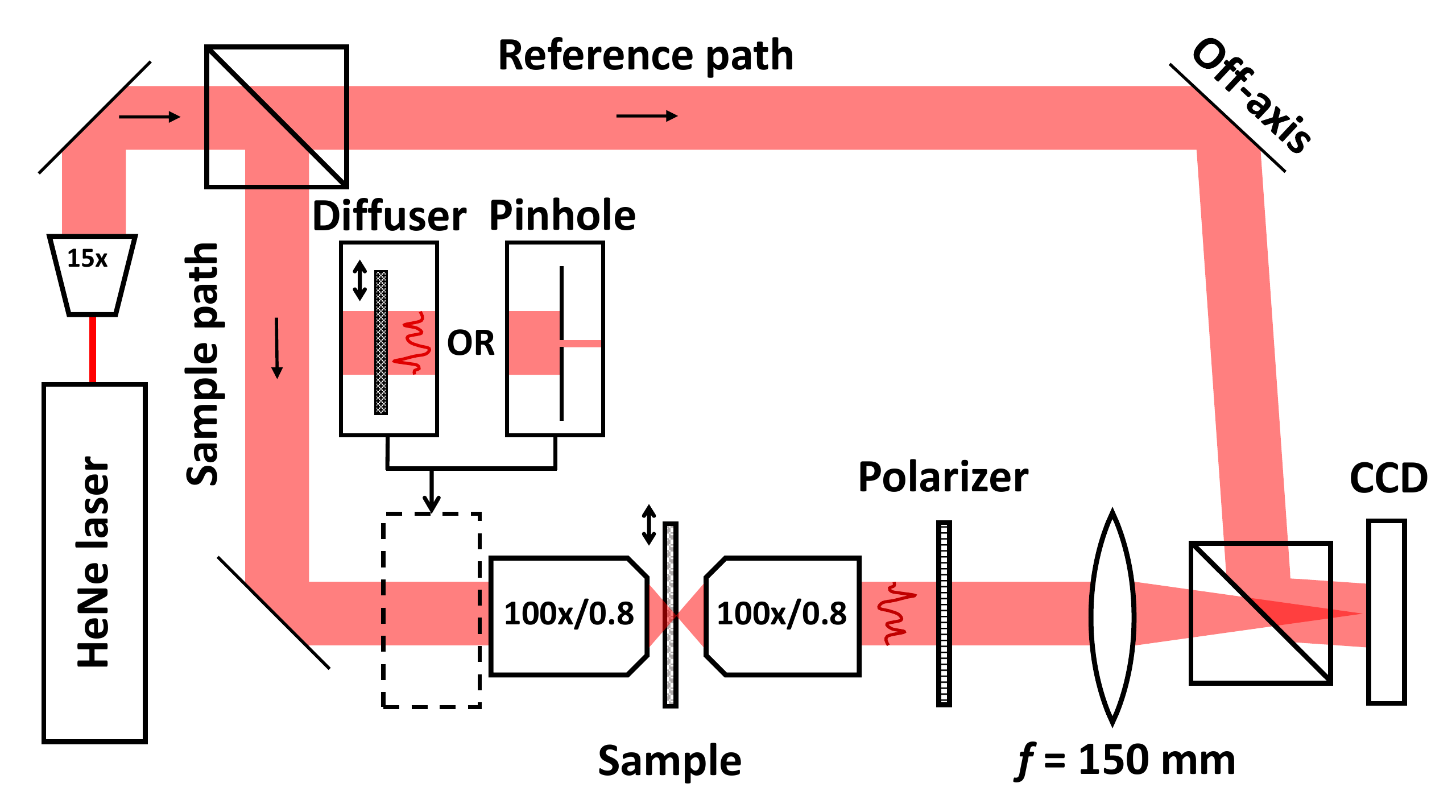}
	\label{fig:setup}
\end{figure*}	
\noindent Schematic of the experimental setup used for measuring both the average light field transmission ($P_W$) and generalized correlation functions ($C_W$). The pinhole and diffuser are used in the $P_W$ and $C_W$ experiments, respectively. Both the diffuser and the sample holder are placed on a translation stage.


\begin{thebibliography}{10}
		\expandafter\ifx\csname url\endcsname\relax
		\def\url#1{\texttt{#1}}\fi
		\expandafter\ifx\csname urlprefix\endcsname\relax\def\urlprefix{URL }\fi
		\providecommand{\bibinfo}[2]{#2}
		\providecommand{\eprint}[2][]{\url{#2}}
		
		\bibitem{Katz:2012}
		\bibinfo{author}{Katz, O.}, \bibinfo{author}{Small, E.} \&
		\bibinfo{author}{Silberberg, Y.}
		\newblock \bibinfo{title}{Looking around corners and through thin turbid layers
			in real time with scattered incoherent light}.
		\newblock \emph{\bibinfo{journal}{Nat. Photon.}} \textbf{\bibinfo{volume}{6}},
		\bibinfo{pages}{549--553} (\bibinfo{year}{2012}).
		
		\bibitem{Bertolotti:2012}
		\bibinfo{author}{Bertolotti, J.} \emph{et~al.}
		\newblock \bibinfo{title}{Non-invasive imaging through opaque scattering
			layers}.
		\newblock \emph{\bibinfo{journal}{Nature}} \textbf{\bibinfo{volume}{491}},
		\bibinfo{pages}{232--234} (\bibinfo{year}{2012}).
		
		\bibitem{Katz:2014}
		\bibinfo{author}{Katz, O.}, \bibinfo{author}{Heidmann, P.},
		\bibinfo{author}{Fink, M.} \& \bibinfo{author}{Gigan, S.}
		\newblock \bibinfo{title}{Non-invasive single-shot imaging through scattering
			layers and around corners via speckle correlations}.
		\newblock \emph{\bibinfo{journal}{Nat. Photon.}} \textbf{\bibinfo{volume}{8}},
		\bibinfo{pages}{784--790} (\bibinfo{year}{2014}).
		
		\bibitem{Yang:2014}
		\bibinfo{author}{Yang, X.}, \bibinfo{author}{Pu, Y.} \&
		\bibinfo{author}{Psaltis, D.}
		\newblock \bibinfo{title}{Imaging blood cells through scattering biological
			tissue using speckle scanning microscopy}.
		\newblock \emph{\bibinfo{journal}{Opt. Express}} \textbf{\bibinfo{volume}{22}},
		\bibinfo{pages}{3405--3413} (\bibinfo{year}{2014}).
		
		\bibitem{Freund:1988}
		\bibinfo{author}{Freund, I.}, \bibinfo{author}{Rosenbluh, M.} \&
		\bibinfo{author}{Feng, S.}
		\newblock \bibinfo{title}{Memory effects in propagation of optical waves
			through disordered media}.
		\newblock \emph{\bibinfo{journal}{Phys. Rev. Lett.}}
		\textbf{\bibinfo{volume}{61}}, \bibinfo{pages}{2328} (\bibinfo{year}{1988}).
		
		\bibitem{Feng:1988}
		\bibinfo{author}{Feng, S.}, \bibinfo{author}{Kane, C.}, \bibinfo{author}{Lee,
			P.~A.} \& \bibinfo{author}{Stone, A.~D.}
		\newblock \bibinfo{title}{Correlations and fluctuations of coherent wave
			transmission through disordered media}.
		\newblock \emph{\bibinfo{journal}{Phys. Rev. Lett.}}
		\textbf{\bibinfo{volume}{61}}, \bibinfo{pages}{834} (\bibinfo{year}{1988}).
		
		\bibitem{Li:1994}
		\bibinfo{author}{Li, J.H.}, \bibinfo{author}{Genack, A. Z.}, 
		\newblock \bibinfo{title}{Correlation in laser speckle}.
		\newblock \emph{\bibinfo{journal}{Phys. Rev. E}}
		\textbf{\bibinfo{volume}{49}}, \bibinfo{pages}{4530--4533} (\bibinfo{year}{1994}).
		
		\bibitem{Schott2015}
		\bibinfo{author}{Schott, S.}, \bibinfo{author}{Bertolotti, J.},
		\bibinfo{author}{L\'{e}ger, J.-F.}, \bibinfo{author}{Bourdieu, L.} \&
		\bibinfo{author}{Gigan, S.}
		\newblock \bibinfo{title}{Characterization of the angular memory effect of
			scattered light in biological tissues}.
		\newblock \emph{\bibinfo{journal}{Opt. Express}} \textbf{\bibinfo{volume}{23}},
		\bibinfo{pages}{13505--13516} (\bibinfo{year}{2015}).
		
		\bibitem{Judkewitz:2015}
		\bibinfo{author}{Judkewitz, B.}, \bibinfo{author}{Horstmeyer, R.},
		\bibinfo{author}{Vellekoop, I.~M.}, \bibinfo{author}{Papadopoulos, I.~N.} \&
		\bibinfo{author}{Yang, C.}
		\newblock \bibinfo{title}{Translation correlations in anisotropically
			scattering media}.
		\newblock \emph{\bibinfo{journal}{Nature Phys.}} \textbf{\bibinfo{volume}{11}},
		\bibinfo{pages}{684--689} (\bibinfo{year}{2015}).
		
		\bibitem{Mertz2015}
		\bibinfo{author}{Mertz, J.}, \bibinfo{author}{Paudel, H.} \&
		\bibinfo{author}{Bifano, T.~G.}
		\newblock \bibinfo{title}{Field of view advantage of conjugate adaptive optics
			in microscopy applications}.
		\newblock \emph{\bibinfo{journal}{Appl. Opt.}} \textbf{\bibinfo{volume}{54}},
		\bibinfo{pages}{3498--3506} (\bibinfo{year}{2015}).
		
		\bibitem{Bastiaans:79}
		\bibinfo{author}{Bastiaans, M.~J.}
		\newblock \bibinfo{title}{Wigner distribution function and its application to
			first-order optics}.
		\newblock \emph{\bibinfo{journal}{J. Opt. Soc. Am.}}
		\textbf{\bibinfo{volume}{69}}, \bibinfo{pages}{1710--1716}
		(\bibinfo{year}{1979}).
		
		\bibitem{Zhang:2009}
		\bibinfo{author}{Zhang, Z.} \& \bibinfo{author}{Levoy, M.}
		\newblock \bibinfo{title}{Wigner distributions and how they relate to the light
			field}.
		\newblock In \emph{\bibinfo{booktitle}{Computational Photography (ICCP), 2009
				IEEE International Conference on}}, \bibinfo{pages}{1--10}
		(\bibinfo{organization}{IEEE}, \bibinfo{year}{2009}).
		
		\bibitem{liu:2015}
		\bibinfo{author}{Liu, H.-Y.} \emph{et~al.}
		\newblock \bibinfo{title}{3d imaging in volumetric scattering media using
			phase-space measurements}.
		\newblock \emph{\bibinfo{journal}{Opt. Express}} \textbf{\bibinfo{volume}{23}},
		\bibinfo{pages}{14461--14471} (\bibinfo{year}{2015}).
		
		\bibitem{Wax1998}
		\bibinfo{author}{Wax, A.} \& \bibinfo{author}{Thomas, E.}
		\newblock \bibinfo{title}{Measurement of smoothed {W}igner phase-space
			distributions for small-angle scattering in a turbid medium}.
		\newblock \emph{\bibinfo{journal}{J. Opt. Soc. Am. A}}
		\textbf{\bibinfo{volume}{15}}, \bibinfo{pages}{1896--1908}
		(\bibinfo{year}{1998}).
		
		\bibitem{cheng2000propagation}
		\bibinfo{author}{Cheng, C.-C.} \& \bibinfo{author}{Raymer, M.}
		\newblock \bibinfo{title}{Propagation of transverse optical coherence in random
			multiple-scattering media}.
		\newblock \emph{\bibinfo{journal}{Phys. Rev. A}} \textbf{\bibinfo{volume}{62}},
		\bibinfo{pages}{023811} (\bibinfo{year}{2000}).
		
		\bibitem{Papadopoulos2016}
		\bibinfo{author}{Papadopoulos, I.}, \bibinfo{author}{Jouhanneau, J.},
		\bibinfo{author}{Poulet, J.} \& \bibinfo{author}{Judkewitz, B.}
		\newblock \bibinfo{title}{Scattering compensation by focus scanning holographic
			aberration probing (f-sharp)}.
		\newblock \emph{\bibinfo{journal}{Nat. Photon.}} \textbf{\bibinfo{volume}{11}},
		\bibinfo{pages}{116--123} (\bibinfo{year}{2017}).
		
		\bibitem{Park2015}
		\bibinfo{author}{Park, J.-H.}, \bibinfo{author}{Sun, W.} \&
		\bibinfo{author}{Cui, M.}
		\newblock \bibinfo{title}{High-resolution in vivo imaging of mouse brain
			through the intact skull}.
		\newblock \emph{\bibinfo{journal}{Proc. Nat. Acad. Sci.}}
		\textbf{\bibinfo{volume}{12}}, \bibinfo{pages}{9236--9241}
		(\bibinfo{year}{2015}).
		
		\bibitem{Hsu:2017}
		\bibinfo{author}{Hsu, C.~W.}, \bibinfo{author}{Liew, S.~F.},
		\bibinfo{author}{Goetschy, A.}, \bibinfo{author}{Cao, H.} \&
		\bibinfo{author}{Stone, A. D.}
		\newblock \bibinfo{title}{Correlation-enhanced control of wave focusing in disordered media}.
		\newblock \emph{\bibinfo{journal}{Nature Phys. Advance online publication}}
		(\bibinfo{year}{2017}).
		
		\bibitem{Tanski:2004}
		\bibinfo{author}{Tanski, I.~A.}
		\newblock \bibinfo{title}{Fundamental solution of Fokker-Planck equation}.
		\newblock \emph{\bibinfo{journal}{arXiv preprint nlin/0407007}}
		(\bibinfo{year}{2004}).
		
	\end{thebibliography}
\end{document}